\documentclass[twocolumn]{elsart3p}

\usepackage{graphics}
\usepackage{epsfig}
\usepackage{amssymb}

\begin{document}

\begin{frontmatter}

\title{The Ginzburg-Landau theory in application}
\author{M. V. Milo\v{s}evi\'{c}\corauthref{cor1}} and
\author{R. Geurts}

\address{Departement Fysica, Universiteit Antwerpen,
Groenenborgerlaan 171, B-2020 Antwerpen, Belgium}

\corauth[cor1]{Corresponding author. \\{\it E-mail}:
milorad.milosevic@ua.ac.be}

\begin{abstract}
A numerical approach to Ginzburg-Landau (GL) theory is demonstrated
and we review its applications to several examples of current
interest in the research on superconductivity. This analysis also
shows the applicability of the two-dimensional approach to thin
superconductors and the re-defined effective GL parameter $\kappa$.
For two-gap superconductors, the conveniently written GL equations
directly show that the magnetic behavior of the sample depends not
just on the GL parameter of two bands, but also on the ratio of
respective coherence lengths.
\end{abstract}

\end{frontmatter}

Whenever a new scientific discovery is made, researchers must strive
to explain it theoretically. In the case of superconductivity, it
took more than two decades after its experimental discovery before
the London theory was developed \cite{London}. However, the London
theory treats vortices as point-like objects and does not take into
account the finite size and the inner structure of the vortex. In
1950, Landau and Ginzburg \cite{GinzburgLandau} developed a
phenomenological theory, which combined Landau's theory of
second-order phase transitions with a Schr\"{o}dinger-like wave
equation. Over the past 50 years, this theory had great success in
explaining macroscopic properties of superconductors (such as the
division of superconductors into two categories now referred to as
type-I and type-II, and the description of the mixed state of
type-II superconductors \cite{Abrikosov}), but it was also immensely
useful for description of mesoscopic superconducting samples
\cite{mesoreview,schw}. Even the full, microscopic Bardeen, Cooper,
and Schrieffer (BCS) theory of superconductivity reduces to the
Ginzburg-Landau (GL) theory close to the critical temperature
\cite{Gorkov}.

The GL theory extends Landau's theory of second-order phase
transitions \cite{Landau} to a spatially varying complex order
parameter $\psi(\vec{r})$ which is nonzero at $T<T_{c}$ and vanishes
at $T\geq T_{c}$ through a second order phase transition. The
resulting gradient term is made gauge-invariant by combining it with
the vector potential $\vec{A}(\vec{r})$ where $\nabla \times
\vec{A}(\vec{r}) =\vec{h}(\vec{r})$ is the local magnetic field.

The two Ginzburg-Landau equations are obtained by minimization of
the GL free energy functional $\mathcal{F}\{\psi, \vec{A}\}$ with
respect to $\psi$ and $\vec{A}$
\begin{eqnarray}
\mathcal{F}\{\psi, \vec{A}\}&=& \frac{H_{c}^{2}}{4\pi}\int \left[
-|\psi|^{2}+\frac{1}{2}|\psi|^{4}+\frac{1}{2}|(-i\nabla-\vec{A})
\psi|^{2} \right. \nonumber \\& &\left.
+\kappa^{2}(\vec{h}-\vec{H}_{0})^{2} \right]dV, \label{GLfunc}
\end{eqnarray}
where $\kappa$ is the GL parameter given as a ratio of magnetic
penetration depth $\lambda$ and the coherence length $\xi$, and
$\vec{H}_{0}$ denotes the applied magnetic field. Eq. (\ref{GLfunc})
is given in dimensionless form, where all distances are measured in
$\xi(T)$, the vector potential $\vec{A}$ in $c\hbar /2e\xi$, the
magnetic field $\vec{H}$ in $H_{c2}=c\hbar /2e\xi ^{2}$, and the
order parameter $\psi$ in $\psi_{0}=\sqrt{-\alpha /\beta}$, with
$\alpha $, $\beta $ being the material dependent coefficients.

Every part of Eq. (\ref{GLfunc}) describes some physical property.
In principle, it is possible to introduce some extra terms in the
energy functional in order to describe the superconducting state
deeper in the superconducting phase (see, for example, Ref.
\cite{Neumann}), but the achieved corrections are very small and are
rarely considered. The first part of Eq. (\ref{GLfunc}) is the
expansion of the energy difference between the superconducting and
normal state for a homogeneous superconductor in the absence of an
applied magnetic field\footnote{It can be shown from the microscopic
theory that only even powers of $|\psi|$ appear in this expansion.}
near the zero-field critical temperature $T_{c0}$. The coefficient
$\alpha$ is negative and changes sign as temperature is increased
over $T_{c0}$ ($\alpha \propto (T-T_{c0})$), while $\beta$ is a
positive constant, independent of temperature. Therefore, the
Cooper-pair density corresponding to temperatures below $T_{c0}$ and
in absence of magnetic field is $|\psi_{0}|^{2}=-\alpha/\beta$.

The next term in Eq. (\ref{GLfunc}) is clearly the kinetic energy of
the Cooper-pairs $ \frac{1}{2m^{\ast}}\left|
\left(-i\hbar\nabla-\frac{2e} {c}\vec{A}\right) \psi\right| ^{2}$,
where $m^{\ast}$ is the effective mass of a Cooper-pair. It
describes the energy cost when the superconducting density is
non-homogeneous.

The last term in Eq. (\ref{GLfunc}) describes the energy of the
magnetic field of the supercurrents, which measures the response of
the superconductor to an external field and is nothing else than the
difference between the local and applied magnetic field. Note that
for a superconductor in an external field, the equilibrium state is
not defined by the Helmholtz free energy but the Gibbs free energy.

The phenomenological GL theory is one of the most elegant and
powerful concepts in physics, which was applied not only to
superconductivity (see textbooks
\cite{Degennes,Tinkham,Schmidt,SaintJames}) but also to other phase
transitions, to nonlinear dynamics, to dissipative systems with
self-organizing pattern formation, and even to cosmology. In what
follows, we describe a numerical approach towards solving the GL
equations, with several particular modifications for special
problems in superconductivity.

\section*{Numerical approach in general}
In what follows, we describe a numerical method used to solve the
Ginzburg-Landau equations, whose solution minimizes the free energy.
Using the dimensionless variables as explained above and the London
gauge, div$\vec {A}=0$, GL equations can be written in the following
form:
\begin{equation}
\left(  -i\nabla-\vec{A}\right) ^{2}\psi =\psi\left( 1-\left|
\psi\right|  ^{2}\right), \label{GLD1}
\end{equation}
\begin{equation}
-\kappa^{2}\Delta\vec{A}=\frac{1}{2i}\left(  \psi^{\ast
}\nabla\psi-\psi\nabla\psi^{\ast}\right) -\left| \psi\right|
^{2}\vec{A}. \label{GLD2}
\end{equation}

The Neumann boundary condition on the sample surfaces takes the form
\begin{equation}
\left.  \vec{n}\cdot\left(  -i\nabla -\vec{A}\right) \psi\right|
_{boundary}=0. \label{GLDB}
\end{equation}

For the fixed applied magnetic field, we solve the two coupled
Ginzburg-Landau equations {\it self-consistently}. Both equations
are solved using the link variable approach \cite{Kato} for a
finite-difference representation of the order parameter and the
vector potential on a uniform cartesian space grid $(x,y)$ with a
typical grid spacing of less than $0.1\xi$.

For a given applied magnetic field, we start from the applied vector
potential as initial condition in our calculation, as if no
superconductor is present. The first step is to solve the first GL
equation (\ref{GLD1}). According to Kato \textit{et al.}
\cite{Kato}, Eq. (\ref{GLD1}) can be written as
\begin{equation}
\frac{\partial\psi}{\partial t}=-\left[  \left(  \frac
{\nabla}{i}-\vec{A}\right) ^{2}\psi+\left( \left| \psi\right|
^{2}-1\right)  \psi\right]  +\widetilde{f} (\vec{r},t),
\label{timekato}
\end{equation}
where time relaxation is included on the left side, and
$\widetilde{f}(\vec{r},t)$ is a dimensionless random force. It is
essential to put the gauge field $A$ on the links of the
computational lattice, which is achieved by introducing the link
variables between $\vec{r}_{1}$ and $\vec{r}_{2}$ as
\begin{equation}
U_{\mu}^{\vec{r}_{1},\vec{r}_{2}}\equiv\exp\left[-i
\int_{\vec{r}_{1}}^{\vec{r}_{2}}\vec{A}_{\mu }\left( \vec{r}\right)
\cdot d\vec{\mu}\right], \label{disklink}
\end{equation}
with $\mu=x,y,z$.

In our calculation, the whole system is mapped on a rectangular
grid. The first term from Eq. (\ref{timekato}) is discretized as
(the index $j$ denotes the lattice point of interest)
\begin{eqnarray}
&&\left(\frac{\nabla_{\mu}}{i}-A_{\mu}\right)^{2}\psi_{j}=-\nabla^{2}_{\mu}\psi_{j}
+i\nabla_{\mu}(A_{\mu}\psi_{j})+\nonumber \\
&&+A_{\mu}^{2}\psi_{j}+iA_{\mu}\nabla_{\mu}\psi_{j}=\frac{1}{U_{\mu}^{j}}\left(
-2iA_{\mu}U_{\mu}^{j}\nabla_{\mu}\psi_{j}\right.\nonumber\\
&&\left.-iU_{\mu}^{j}\psi_{j}(\nabla_{\mu}A_{\mu}-iA_{\mu}^{2})
+U_{\mu}^{j}\nabla_{\mu}^{2}\psi_{j} \right). \label{kato_discr}
\end{eqnarray}
After substituting $\nabla_{\mu}U_{\mu}^{j}=-iA_{\mu}U_{\mu}^{j}$
and
$\nabla_{\mu}^{2}U_{\mu}^{j}=-iU_{\mu}^{j}(\nabla_{\mu}A_{\mu}-iA_{\mu}^{2})$,
and some trivial transformations, we obtain
\begin{equation}
\left(\frac{\nabla_{\mu}}{i}-A_{\mu}\right)^{2}\psi_{j}=\frac{1}{U_{\mu}^{j}}\nabla_{\mu}(\nabla_{\mu}(U_{\mu}^{j}\psi_{j})).
\end{equation}
Finally, for $\mu=x$ (analogously for $\mu=y,z$) we have
\begin{eqnarray}
&&\left(\frac{\nabla_{x}}{i}-A_{x}\right)^{2}\psi_{j}=\nonumber\\&=&\frac{1}{U_{x}^{j}}\frac{1}{a_{x}}\left(
\frac{U_{x}^{j+1}\psi_{k}-U_{x}^{j}\psi_{j}}{a_{x}}-\frac{U_{x}^{j}\psi_{j}-U_{x}^{j-1}\psi_{j-1}}{a_{x}}\right)\nonumber\\
&=&\frac{U_{x}^{j+1,j}\psi_{j+1}-2\psi_{j}+U_{x}^{j-1,j}\psi_{j-1}}{a_{x}^{2}}.
\end{eqnarray}
The discretized Ginzburg-Landau equation can be now written in full
as
\begin{eqnarray}
\frac{\partial\psi}{\partial
t}&=&\frac{U^{kj}_{x}\psi_{k}}{a_{x}^{2}}+\frac{U^{ij}_{x}\psi_{i}}
{a_{x}^{2}}+\frac{U^{mj}_{y}\psi_{m}}{a_{y}^{2}}+\frac{U^{nj}_{y}\psi_{n}}{a_{y}^{2}}\nonumber\\
&+&\frac{U^{gj}_{y}\psi_{g}}{a_{z}^{2}}+\frac{U^{hj}_{y}\psi_{h}}{a_{z}^{2}}
-2\psi_{j}\left(\frac{1}{a_{x}^{2}}+\frac{1}{a_{y}^{2}}+\frac{1}{a_{z}^{2}}\right)
\nonumber\\&-&\left( \left| \psi_{j}\right|  ^{2}-1\right) \psi_{j}
+\widetilde{f}_{j}(t), \label{kato_car}
\end{eqnarray}
where different indices denote adjacent grid points to point $j$
along three axis. In general, this approach works for $a_{x}\neq
a_{y}\neq a_{z}$.

Using Eq. (\ref{kato_car}), with chosen initial vector potential, we
solve for the value of the order parameter $\psi$ in every grid
point. These values we can implement to calculate the local current
densities $j_{x,y,z}$. The right side of Eq. (\ref{GLD2}) can be
written as
\begin{equation}
\vec{j}=\frac{1}{2}\left[ \psi^{\ast}\left(
\frac{1}{i}\nabla-\vec{A}\right) \psi+\psi\left(
\frac{1}{i}\nabla-\vec{A}\right) ^{\ast}\psi^{\ast}\right],
\end{equation}
where again link variable approach comes into play through similar
transformations as in Eqs. (\ref{kato_discr})-(\ref{kato_car})
\begin{eqnarray}
&&\left(  \frac{1}{i}\nabla_{x}-A_{x}\right)
\psi_{j}\rightarrow\nonumber\\&&-i\frac{1}{U_{x}^{j}}\nabla_{x}(U_{x}^{j}\psi_{j})
=-i\frac{U_{x}^{kj}\psi_{k}-\psi_{j}}{a_{x}}. \label{disccur}
\end{eqnarray}

From the supercurrents a new value for the vector potential can be
calculated using the second GL equation. This is a Poisson-type of
equation, which we solve using a Fourier transformation. The then
obtained vector potential is used (in part, typically $5\%$) to
update the current vector potential gradually. The updated vector
potential is substituted back in the first GL equation and the whole
procedure is repeated until a convergent solution of both GL
equations is found.

\subsection*{Temperature dependence}
The temperature dependence of characteristic lengths $\xi$ and
$\lambda$ and the critical field is incorporated in GL theory as
$\xi(T)=\frac{\xi(0)}{\sqrt{\left|  1-T/T_{c0}\right| }}$,
$\lambda(T)=\frac{\lambda(0)}{\sqrt{\left| 1-T/T_{c0}\right| }}$,
and $H_{c2}(T) =H_{c2}(0)\left|  1-\frac{T}{T_{c0}}\right|$.
However, if GL equations are scaled to temperature dependent units
as given above, that allows us to consider any temperature
dependence of the parameters prior to the start of the simulation. A
number of works in the past were dedicated to theoretically improved
temperature dependence of $\xi$, $\lambda$, and $\kappa$. Underlying
effects for reported differences may be nonlocality, clean/dirty
limit effects, strong coupling. GL parameter $\kappa$ is temperature
independent in GL theory, but for e.g. type-I superconductors
Ginzburg himself suggested the correction
$\kappa(T)=\kappa(0)/(1+t^2)$ (the two-fluid model), Bardeen gave
$\kappa(T)=\kappa(0)/\sqrt{1+t^2}$, while Gorkov found
$\kappa(T)=\kappa(0)(1-0.24t^2+0.04t^4)$, where $t=T/T_c0$. Although
these formulae go beyond GL theory, we emphasize here that they can
be implicitly included in the calculation, with aim to improve the
overall validity of the theory.

It is also worth mentioning that simulations can account for the
exact experimental procedure with respect to cooling. For the
zero-field cooled regime the initial value of $\psi$ should be taken
$\approx 1$, and in the field-cooled regime opposite, i.e.
$\psi\approx 0$ since superconductivity nucleates from the normal
state.

\section*{Fourier transform and 2D approximation}
It is well known that in superconductors the magnetic field
penetrates only into a relatively small depth $\lambda$ and that
screening currents flow in the surface layer, decaying exponentially
in the bulk beyond this length. However, one question arises: what
happens when the superconductor is thinner than the London
penetration depth? Tinkham argued, as is now generally accepted,
that the currents in thin superconductors may be considered constant
over the thickness. Consequently, they have {\it no} $z$-component,
and the boundary condition (\ref{GLDB}) is automatically fulfilled
at top and bottom surfaces of the sample. Prozorov {\it et al.}
\cite{prozorov} confirmed Tinkham's original assumption, considering
in detail the current density (in)homogeneity throughout the
thickness of superconducting films.

Therefore, for a superconductor with thickness $d<\lambda,\xi$ we
assume the uniform distribution of current in the $z$ direction.
From the first GL equation (\ref{GLD1}) then follows the same
behavior for the order parameter, and the 3D problem is reduced to a
2-dimensional superconductor. However, an issue of solving the GL
equation for the vector potential remains, and here we discuss the
details.

We are actually solving three equations of shape $-\kappa^2 \Delta A
= j$. From Fourier theory we know that
\begin{eqnarray}
 F(k) & = & \int_{-\infty}^{\infty} f(x) \exp(-2\pi i x k ) dx,  \\
 f(x) & = & \int_{-\infty}^{\infty} F(k) \exp(2\pi i x k ) dx.
\end{eqnarray}
We use this to solve the equation analytically in the $z$-direction.
In the $x$- and $y$-direction, we solve numerically using FFT which
is based on the following Fourier relationship
\begin{eqnarray}
 b_n & = & \frac{2}{N} \sum_{j=1}^{N-1} f_j \sin(j\pi \frac{n}{N}), \\
 f_j & = & \sum_{n=1}^{N-1} b_n \sin(j\pi \frac{n}{N}),
\end{eqnarray}
where we have chosen the sine transform, to respect the boundary
condition $\vec{A}(r\rightarrow \infty)\rightarrow 0$\footnote{The
$\vec{A}$ we solve for here represents solely the field induced by
the superconductor and does not include the applied field.}.

Every component of the vector potential can therefore be represented
as
\begin{eqnarray}
&&A(x,y,z) = \sum_{i=1}^{N-1} \sum_{j=1}^{N-1} \nonumber\\
&&\int_{-\infty}^{\infty} {dk} a_{ij}(k) \exp(2\pi i z k)
\sin(\frac{i\pi x}{N}) \sin(\frac{j\pi y}{N}),
\end{eqnarray}
where $x$ and $y$ are integer numbers between $1$ and $N$ and $z$
can be any real number. We assume the sample with thickness $d$ to
be centered at $z=0$. The Laplacian of $A$ can be obtained
analytically as:
\begin{eqnarray} \Delta A(x,y,z)
= \sum_{i=1}^{N-1} \sum_{j=1}^{N-1} \int_{-\infty}^{\infty} {dk}
\left(-(2\pi k)^2 - q^2 \right)  & & \nonumber\\ a_{ij}(k) \exp(2\pi
i z k) \sin(\frac{i\pi x}{N}) \sin(\frac{j\pi y}{N}), & &
\end{eqnarray}
where we introduced $q^2 \equiv (i \pi/L_x)^2 + (j \pi/L_y)^2$.
$L_{x,y}$ is the size of the simulation region in the $x$- and
$y$-direction respectively.

Now we introduce $j(x,y)$ as the current uniform over the sample
thickness
\[ j(x,y,z) = j(x,y) \Pi(z, -d/2, d/2), \]
where $\Pi$ represents the step-like function which is $1$ inside
the interval $[-d/2, d/2]$ and $0$ outside. The Fourier transform of
$\Pi$ yields:
\[ \Pi(k) = d \frac{\sin(\pi k d)}{\pi k d}.  \]
Using sine transform, we can express the current as:
\[ j(x,y) = \sum_{i=1}^{N-1} \sum_{j=1}^{N-1} b_{ij} \sin(i \pi x/ N) \sin(j \pi y/ N), \]
with coefficients:
\[ b_{ij}  \equiv \frac{4}{N^2} \sum_{x=1}^{N-1} \sum_{y=1}^{N-1}
j(x,y) \sin(i\pi \frac{x}{N}) \sin(j\pi \frac{y}{N}). \]

Substituting this into initial GL equation we get:
\begin{eqnarray}
\sum_{i=1}^{N-1} \sum_{j=1}^{N-1}\sin(\frac{i\pi x}{N})
\sin(\frac{j\pi y}{N}) \int_{-\infty}^{\infty} {dk} \exp(2\pi i z k)
& & \nonumber \\  \left(  \kappa^2 [(2\pi k)^2 + q^2 ] a_{ij}(k) -
b_{ij}d \frac{\sin(\pi k d)}{\pi k d} \right) = 0, & &
\end{eqnarray}
which can only be true if the terms in the brackets equal zero, i.e.
when
\[
\frac{\kappa^2}{d}  a_{ij}(k) =   \frac{b_{ij}}{(2\pi k)^2 + q^2
}\frac{\sin(\pi k d)}{\pi k d}.
\]
We now revert to the definition of $A$ in terms of its Fourier
transform, and obtain:
\begin{eqnarray}
\frac{\kappa^2}{d}&&A(x,y,z)= \sum_{i=1}^{N-1} \sin(i \pi x/ N)
\sum_{j=1}^{N-1} \sin(j \pi y/ N) b_{ij} \nonumber \\
&&\int_{-\infty}^{\infty} {dk} \exp(2\pi i z k) \frac{1}{(2\pi k)^2
+ q^2 }\frac{\sin(\pi k d)}{\pi k d}.
\end{eqnarray}
The last term can be integrated analytically to get:
\begin{eqnarray}
 \frac{1}{d q^2} \left( 1 - \cosh(qz) \exp(-d q/2) \right), & & \textrm{if  }  z < d/2,  \nonumber\\
 \frac{1}{d q^2} \left( \sinh(d q/2) \exp(-q z) \right), &  & \textrm{if  } z >
 d/2.\nonumber
\label{eq1}
\end{eqnarray}
We insert $z=0$ directly in above equation, and obtain $\frac{1}{d
q^2} \left( 1 - \exp(-d q/2) \right)$.

Recapitulating, we can express $A(x,y, z=0)$ in its Fourier
components:
\begin{eqnarray}
&&A(x,y,0)=\nonumber\\ &&\sum_{i=1}^{N-1} \sin(i \pi x/ N)
\sum_{j=1}^{N-1} \sin(j \pi y/ N) a_{ij},
\end{eqnarray}
with as
coefficients:
\begin{equation} a_{ij} = \frac{b_{ij}}{\kappa^2}
\frac{ 1 -\exp(-d q/2)}{q^2}.
\end{equation}
Note that the common definition for effective GL parameter for thin
superconductors $\kappa_{eff} = \kappa^2 / d$ is therefore just a
{\it limiting case} obtained for extremely thin samples
($j(x,y,z)=\delta(z)j(x,y)$).

Alternatively, one can calculate the {\it mean} vector potential
(averaged in $z$-direction) inside the sample, instead of the values
in $z=0$ plane only. In that case, the Fourier coefficients become:
\[ <a_{ij}>_z = \frac{1}{d}\frac{b_{ij}}{\kappa^2 q^2} \left( d - 2\frac{\sinh(dq/2)}q \right)  \exp(-d ij/2).\]

\section*{The anisotropy}
Modern fabrication methods enable experiments on various hybrid
structures. Among others, one can think of complex 3D
superconductor-metal hybrids, as well as samples made of different
superconducting materials. In addition, materials such as high-$T_c$
superconductors are mostly layered, and exhibit clear anisotropy in
different directions. In what follows, we show how those are
incorporated in our GL formalism.

To begin with, we consider a superconductor-metal hybrid. As already
pointed out by deGennes, the leakage of Cooper-pairs from the
superconductor into the metal can be modeled through the modified
boundary condition for a superconductor-normal metal interface as
\begin{equation}
\left.  \vec{n}\cdot(-i\hbar\nabla-\frac{2e^{\ast}}
{c}\vec{A})\psi\right|_{interface}=\frac{i}{b}\psi, \label{GLBG}
\end{equation}
where quantity $b$ is positive and measures the distance outside the
boundary (in the normal metal) where the order parameter becomes
zero if the slope at the interface is maintained. As a consequence,
the discretized first GL equation (\ref{kato_car}) must be modified
at the boundary of the superconductor, and the terms
$\psi_j/a_{\mu}^2$ perpendicular to the interface get a prefactor
$(1-b)$.

In the case of a superconductor-superconductor hybrid, the situation
is far more complex. The simplest scenario is that only $T_c$ is
different between two materials, in which case the only parameter
$\alpha$ varies in the sample. This is put in GL equations directly,
and generates no extra terms in the discretization. Note that the
crossing currents are also calculated directly, and no special
boundary condition (such as Eq. (\ref{GLBG})) is needed. In the case
of varied mean free path, both $\alpha$ and $\beta$ change
(proportional to $l$ and $l^2$ respectively), but this still does
not generate additional terms in GL equations and can be put in the
calculations directly. However, to consider mass anisotropy in the
sample, one must go back to the very derivation of GL equations to
establish that an additional term $\nabla \frac{1}{m(x,y,z)}\left(
-i\nabla-\vec{A}\right)\psi$ appears on the right side of Eq.
(\ref{GLD1}). This term is then discretized using Eq.
(\ref{disccur}) and added to Eq. (\ref{kato_car}) in the numerical
procedure.

Note that for real layered samples (such as high-$T_c$ ones) the
coupling between the layers must also be introduced into GL
equations. This is the known Lawrence-Doniach extension to the GL
model, where Josephson coupling is active between the
superconducting layers separated by an insulator of thickness $d_i$.
As a result, at planes $z=\pm d_i/2$ a term $[\psi_{\pm d_i/2}-
\psi_{\mp d_i/2}\exp(\pm iA_{int}^{z})]/d_i$ is added to the first
GL equation (here $A_{int}^{z}=\int_{-d_i/2}^{d_i/2}A_{z}dz$). The
second GL equation remains the same, but an additional equation
describing the Josephson current enters the calculation at all S-I
boundaries as
\begin{equation}
\label{eq2} j_{z}=\frac{i}{2d_i}\left[\psi_{d_i/2} e^{-iA_{int}^{z}}
\psi_{-d_i/2}^*-\psi_{d_i/2}^* e^{iA_{int}^{z}}
\psi_{-d_i/2}\right].
\end{equation}

\section*{Two-band superconductivity}
Multi-band superconductivity has stirred great interest in the
scientific community following the discovery and further studies of
MgB$_{2}$ \cite{Akimitsu}, having the highest critical temperature
for a non-copper-oxide and non-fullerene superconductor. This
material and some others such as boro-carbides can be described by
two superconducting order parameters (hence the name `two-band' or
`two-gap' superconductor). However the real boom in two- and
multi-band superconductivity followed the last year's discovery of
iron pnictides \cite{hirono}. The concentration of different dopants
and applying of external pressure can tune the electronic, magnetic
and structural properties of these materials. Provided that their
critical temperature is increased above liquid air, the impact of
these materials on technology is already seen as a `new iron age'.

In cases when microscopic structure of the material is not relevant,
the Ginzburg-Landau theory can describe two-gap superconductors. The
first GL equation, for each of the two order parameters now reads:
\begin{eqnarray}
\label{2gapGL1} (-i\nabla - \vec{A})^2 \psi_1 - (\chi_1 -
|\psi_1|^2) \psi_1 -
\frac{\gamma}{\delta} \psi_2  \nonumber\\- \frac{\eta}{\delta} \left( -i \nabla - \vec{A}\right)^2 \psi_{2} = 0 \\
\frac{1}{\alpha}(-i\nabla - \vec{A})^2 \psi_2 - \left(\chi_2 -
|\psi_2|^2 \right) \psi_2  - \frac{\gamma\delta}{m\alpha}\psi_1
\nonumber\\- \frac{\eta\delta}{m\alpha} \left( -i \nabla -
\vec{A}\right)^2 \psi_{1} = 0,
\end{eqnarray}
where $\chi_i=1-T/T_{ci}$ ($T_{ci}$ being the critical temperature
of gap i). $\alpha=(\xi_{10}/\xi_{20})^2$ and
$\delta=\psi_{10}/\psi_{20}=\sqrt{\frac{\alpha_{10}\beta_2}{\beta_1\alpha_{20}}}$,
all measured at $T=0$ and in absence of magnetic field and coupling.
$\gamma$ and $\eta$ are the coupling constants representing
Josephson and drag effect respectively. $m=m_1/m_2$ is the ratio of
the effective masses in two gaps. Lengths are expressed in
$\xi_{10}$, and the order parameters in $\psi_{i0}$.

The second GL equation is a single equation, with contribution from
both condensates. We write it in the following form using the
relation $\frac{\kappa_{1}^{2}}{\kappa_{2}^{2}} = \frac{m}{\delta^2
\alpha}$:
\begin{eqnarray}
-\Delta \vec{A} = & & \frac{1}{\kappa_{1}^{2}}\mathcal{R}\left[
\psi_{1}^{*}\left( -i\nabla - \vec{A}\right)\psi_{1} \right] \nonumber\\
& + & \frac{\alpha}{\kappa_{2}^{2}} \mathcal{R}\left[
\psi_{2}^{*}\left( -i\nabla -
\vec{A}\right)\psi_{2} \right] \nonumber\\
& + & \eta \sqrt{\frac{\alpha}{m}} \frac{1}{\kappa_1\kappa_2}
\mathcal{R}\left[ \psi_{1}^{*}\left( -i\nabla
- \vec{A}\right)\psi_{2} \right] \nonumber\\
& + & \eta \sqrt{\frac{\alpha}{m}} \frac{1}{\kappa_1\kappa_2}
\mathcal{R}\left[ \psi_{2}^{*}\left( - i\nabla -
\vec{A}\right)\psi_{1} \right], \label{2gapGL2}
\end{eqnarray}
where we singled out the GL parameters for both condensates,
although this is not realistic. However, the latter equation makes
directly visible that the magnetic behavior of the sample will
depend not only on $\kappa_1$ and $\kappa_2$, but also on the ratio
of coherence lengths in two condensates (parameter $\alpha$), and
this is of direct relevance to the recently observed type-1.5
superconductivity \cite{moshPRL}. Note also that all terms in Eqs.
(\ref{2gapGL1}-\ref{2gapGL2}) were already present in the same shape
in the single-gap GL equations. Therefore, Eqs.
(\ref{2gapGL1}-\ref{2gapGL2}) can be discretized and solved using
the same numerical scheme as given in preceding sections.

In summary, we have explained in more detail the numerical approach
that discretizes Ginzburg-Landau equations on a three dimensional
Cartesian grid. We further reviewed its applications on
superconducting samples of different topologies, the hybrid systems,
anisotropic samples, layered superconductors, and two-gap
superconductors. For thin samples, we show the modification of the
approach that allows fast solving for order parameter in the 2D
plane, while the vector potential is generally solved in all three
dimensions. We therefore hope to have presented a brief manual which
will be of use for young researchers in the years to come.

\section*{Acknowledgments}
This work was supported by the Flemish Science Foundation (FWO-Vl),
the Belgian Science Policy, and the JSPS/ESF-NES network.


\begin{thebibliography}{00}
\bibitem {London} F. London and H. London, Proc. Roy. Soc. \textbf{A149}, 71 (1935).
\bibitem {GinzburgLandau} V. L. Ginzburg and L. D. Landau, Zh. Eksp. Teor. Fiz. \textbf{20}, 1064 (1950).
\bibitem {Abrikosov} A. A. Abrikosov, Sov. Phys. JETP \textbf{5}, 1175 (1957).
\bibitem{mesoreview} V. V. Moshchalkov, in \textit{Handbook of Nanostructured Materials and Nanotechnology},
pp. 451-525 (Academic Press, New York, 1999).
\bibitem{schw} V. A. Schweigert and F. M. Peeters, Phys. Rev. B {\bf 57},
13817 (1998); V. A. Schweigert, F. M. Peeters, and P. Singha Deo,
Phys. Rev. Lett. \textbf{81}, 2783 (1998).
\bibitem {Gorkov}L. P. Gor'kov, Sov. Phys. JETP \textbf{9}, 1364 (1959).
\bibitem {Landau}L. D. Landau and E. M. Lifshitz, \textit{Statistical
Physics}, 3rd edn., part 1 (Pergamon Press, Oxford, 1980).
\bibitem {Neumann}L. Neumann and L. Tewordt, Z. Phys. \textbf{189}, 55 (1966).
\bibitem {Degennes}P. G. de Gennes, \textit{Superconducting of Metals and
Alloys} (Benjamin, New York, 1966).
\bibitem {Tinkham}M. Tinkham, \textit{Introduction to Superconductivity}
(McGraw Hill, New York, 1975).
\bibitem {Schmidt}V. V. Schmidt, P. M\"{u}ller, A. V. Ustinov, \textit{The
Physics of Superconductors: Introduction to Fundamentals and
Applications} (Springer-Verlag, Berlin Heidelberg, 1997).
\bibitem {SaintJames}D. Saint-James, G. Sarma, and E. J. Thomas, \textit{Type
II Superconductivity} (Pergamon Press, Oxford, 1969).
\bibitem {Kato} R. Kato, Y. Enomoto, and S. Maekawa, Phys. Rev. B \textbf{47},
8016 (1993).
\bibitem{prozorov} R. Prozorov, E. B. Sonin, E. Sheriff, A. Shaulov, and Y.
Yeshurun, Phys. Rev. B {\bf 57}, 13845 (1998).
\bibitem{Akimitsu} J. Nagamatsu, N. Nakagawa, T. Muranaka, Y. Zenitani, and J.
Akimitsu, Nature (London) {\bf 410}, 63 (2001).
\bibitem{hirono} Y. Kamihara, T. Watanabe, M. Hirono, and H. Hosono, J. Am. Chem. Soc. {\bf 128}, 10012 (2006); ibid {\bf 130}, 3296 (2008).
\bibitem{moshPRL} V. V. Moshchalkov, M. Menghini, T. Nishio, Q. H. Chen, A. V. Silhanek, V. H. Dao, L. F. Chibotaru, N. D. Zhigadlo, and J. Karpinski, Phys. Rev.
Lett. {\bf 102}, 117001 (2008).
\end{thebibliography}
\end{document}